# Rocky planet rotation, thermal tide resonances, and the influence of biological activity

Caleb Scharf, Columbia Astrobiology Center & Dept. of Astronomy, Columbia University, MC 5246, 550 West 120th St., New York, NY 10027, USA

**Abstract**

It has been established theoretically that atmospheric thermal tides on rocky planets can lead to significant modifications of rotational evolution, both close to synchronous rotation and at faster rotations if certain resonant conditions are met. Here it is demonstrated that the normally considered dissipative gravitational tidal evolution of rocky planet rotation could, in principle, be 'stalled' by thermal tide resonances for Earth-analog worlds in the liquid water orbital zone of stars more massive than $\sim 0.3 M_\odot$. The possibility of feedback effects between a planetary biosphere and the planetary rotational evolution are examined. Building on earlier studies, it is suggested that atmospheric oxygenation, and ozone production could play a key role in planetary rotation evolution, and therefore represents a surprising but potent form of biological imprint on astronomically accessible planetary characteristics.

Keywords: exoplanet rotation --- thermal tides --- ozone --- biosphere feedback

# 1. Introduction

Differential, inelastic deformations of planets arise due to the gravitational effect of a perturbing body. These deformations can lead to the dissipation of energy in a system and gravitational torques that can accelerate or decelerate orbital motions and rotations. Planetary atmospheres are also subject to thermal tides due to stellar radiation, and these tides can alter atmospheric mass distributions leading to additional torques that are transferred to the solid body motion.

The effect of thermal tides in planetary atmospheres on rotation have been long discussed; including the work of Kelvin (1882) (see also Lindzen & Chapman (1969), Goldreich & Soter (1969), and Ingersoll & Dobrovolskis (1978) in discussing Venus, and more recently for both short-period gas giants (Arras & Socrates 2010) and for near-synchronous rocky exoplanets (Cunha *et al.* 2015, Leconte *et al.* 2015)). Diurnal (weak) and semi-diurnal (strong) tide pressure-wave components are generated (e.g. Lindzen & Chapman 1971), but it is the latter which can give rise to torques acting on planetary rotation through atmosphere-surface friction. For the modern Earth the semi-diurnal tide can be thought of as the movement of atmospheric mass away from the hottest point, which occurs a few hours after midday. Mass then accumulates in bulges on the dayside (approaching the substellar point) and nightside (receding from substellar point). Therefore, the solar torque on the Earth's dayside bulge acts to accelerate rotation, but this torque is currently far smaller than lunar and solar solid body and oceanic torques.

For planetary systems surrounding low-mass stars, the surface-liquid-water orbital zone (also commonly known, albeit misleadingly, as the habitable zone) typically spans a range where star-planet tidal interactions are expected to bring planets' rotation into near synchronous, slow-rotating states in a fraction of the system age. Dissipation during this process will also tend to circularize the orbits (but not reduce the semi-major axis). In these conditions the solid body and oceanic torques become small, and potentially comparable to thermal tide torques.

Consequently, as indicated by Correria & Lasker (2010), and further investigated by Cunha *et al.* (2015) and Leconte *et al.* (2015), torques due to atmospheric thermal tides may actually prevent a planet's rotation from entering a precisely synchronous state, even for rocky worlds with thin, Earth-analog atmospheres. Near synchronous, stable states may include both slow retrograde and slow prograde rotations.

However, much earlier in the rotational evolution of these planets the presence of a resonance state between the frequency of free oscillation of the atmosphere (Lamb 1932) and the semi-diurnal frequency component of thermal tides could also disrupt the expected planetary spin-down. For Earth this resonance has been predicted to occur at faster, historical rotation periods (Zahnle & Walker 1987). Noting that the eigenvalues of horizontal atmospheric thermal tidal wave structure ($\beta = \frac{4R_\oplus^2 \Omega_\oplus^2}{gh}$, where $R_\oplus$ and $\Omega_\oplus$ are planetary radius and rotational angular velocity) are modulated by $h$, the equivalent depth of the wave, with a current value of 7.852 km, the resonant daylength corresponds to $d = \frac{7.852(24)^{\frac{1}{2}}}{h}\ hr$ where $h$ is in km.

To a first approximation the equivalent depth, $h$, of an atmosphere is of the order of the ratio of specific heats ($\gamma$) times the average atmospheric scale height. Therefore, the resonance should scale according to $R\Omega \propto \sqrt{kT/m}$ where $m$ is the mean molecular weight of atmospheric gases, i.e. the resonant daylength scales with $\sqrt{m/T}$.

For the modern Earth the free oscillation equivalent depth is $\sim 10\ km$, therefore resonance must have occurred when $d \sim 21\ hr$. It should be noted that at faster rotations the phase lag angle is such that thermal tidal torque acts to decelerate the rotation, but in crossing resonance the sign of this torque changes and it acts in a prograde sense, giving rise to potential torque balance with lunar and solar torques. Fig 1. illustrates this schematically.

More detailed analyses lead to the observation (Zahnle & Walker 1987) that during the Precambrian on Earth (prior to 600 Myr ago) resonance could have been met, causing a significant enhancement in semi-diurnal atmospheric torques (with surface pressure oscillations perhaps jumping from less than 1 $mbar$ to $\sim 20\ mbar$). The resulting thermal tide torques would have been comparable to the dominant lunar tidal torque at that time (Fig. 1), leading to a 'stalling' of Earth's rotational slow-down while the lunar orbit continued to evolve.

The details of this resonant trapping hinge on numerous factors, including the resonance width (and hence the timescale over which restoring forces must act: a few 10 million to 100 million years), atmospheric composition (through the $\sqrt{m/T}$ scaling and the efficiency of radiation absorption), surface temperature, and specifics of dissipative processes in atmospheric waves. More recent work (Bartlett & Stevenson 2016) has investigated the effect of thermal noise (e.g. Milankovitch-type cycles on $10^3$-$10^4$ year timescales) and concludes that the resonant trap for Earth (considering the lunar tides) could have lasted up to 1000 Myr and could be broken following severe ice ages, where global temperatures might climb rapidly (e.g. $\Delta T \sim 20K$) over $10^7$ years. Higher temperature lowers the daylength for resonance (i.e. moving curve in Fig.1 to the right, by altering atmospheric column height according to the $\sqrt{m/T}$ dependency), allowing the planet to escape back to its long-term spin-down driven by lunar and solar tidal torques. Other mechanisms, such as changes in oceanic dissipation, could also be at play.

The question addressed in this present work is whether Earth-analog exoplanets (matching in approximate mass, composition, temperature, and atmospheric conditions) in the liquid-surface-water zone of lower mass stars could also experience resonant trapping. Since these orbital zones are significantly closer to the parent star than the Earth-Sun system, the possibility exists for stellar tidal torques (i.e. in the absence of a lunar torque) to be large enough to balance the expected thermal tidal torques close to resonance (see below).

Observational constraints on initial planet rotation rates do not yet exist beyond the solar system. However, some theoretical work suggests that terrestrial-type planet populations can have high initial spins, possibly up to the critical angular velocity for rotational instability (Kokubo & Ida 2007). Other studies suggest a broader range of rotation periods is possible, from $10\ hr$ to $10,000\ hr$ (e.g. Miguel & Brunini 2010), although the distribution is relatively flat, implying a significant population of objects in the $10\ hr$ regime. It is assumed here that many Earth-analog worlds have initial rotation rates fast enough that thermal tides will eventually encounter the resonant state as solar tides slow the rotation.

## 2. Results: effects of thermal tide resonance on rocky exoplanets

The potential for resonant thermal tides to stall planetary rotation evolution can be evaluated by considering the orbit-averaged amplitudes of thermal tide torques and stellar tidal torques that can act in opposition. Here the following expression for resonant thermal tide torque is used, assuming zero obliquity and a circular orbit (e.g. Zahnle & Walker 1987):

$$T_{thermal} = \frac{3\pi G M_* R_p^4}{4 a_p^3} \frac{\delta p(a_p)}{g} \sin(2\Delta\varphi)$$

where stellar mass is $M_*$, planetary radius, semi-major axis (circular orbit), and surface gravitational acceleration are $R_p$, $a_p$, $g$ respectively. $\Delta\varphi$ corresponds to the tide phase lag at the planetary surface (about 158 degrees for the modern Earth). It is assumed that the dissipative transfer of this torque into the solid body is fast.

The amplitude of the semi-diurnal pressure fluctuation at resonance is denoted by $\delta p$. A value for $\delta p$ was estimated by Zahnle & Walker (1987) to be close to $\sim 20\ mbar$ for the Earth. In reality this amplitude must be proportional to the amplitude of the semi-diurnal *forcing*, namely the absorption of stellar radiation at the most appropriate altitudes. In order to account for the variation in stellar input of different Earth-analogs in the following calculations a first approximation is made by assuming that $\delta p \propto a_p^{-2}$.

Following Goldreich & Soter (1966), the nominal star-planet gravitational torque is:

$$T_{grav} = \frac{9}{4} \frac{GM_*^2 R_p^5}{Q'_p a_p^6}$$

where $1/Q_p'$ is the effective tidal dissipation function, $Q'_p = Q(1 + \frac{19\mu}{2g\rho R_p})$. Rigidity $\mu$ is assumed to be $\sim 10^{11}$ dyn/cm$^2$ for Earth-analog compositions, and rocky planet density is assumed to be $\sim 5.5$ g/cm$^3$. The parameter $Q$ approximates many complex, not very well understood phenomena, but is often considered to be of the order $\sim 100$ for the modern Earth from empirical arguments.

In Fig. 2 the maximum thermal tide torques at resonance are compared to the opposing stellar tidal torques for an Earth-mass planet occupying the approximate liquid-surface-water orbital range (c.f. habitable zone) for stars of masses 0.1, 0.3, 0.5, 0.7, and 1.0 $M_\odot$. By selecting the potential planet surface-temperature range, the effect of the $\sqrt{m/T}$ dependency of resonance period is somewhat constrained in this study (matching the range considered in Zahnle & Walker 1987).

The orbital distance dependency of the thermal pressure wave amplitude near resonance is accounted for as follows. The orbital radius/semi-major axis at which the stellar flux is equivalent to that at Earth is computed according to $a_{p-equiv} = \sqrt{L_* a_\oplus^2 / L_\odot}$, where stellar luminosities $L_*$ are estimated using the usual main-sequence mass-luminosity scaling laws: $L_*/L_\odot = (M_*/M_\odot)^4$ for stellar masses between 0.43 and 2 solar masses, and $L_*/L_\odot \approx 0.23 (M_*/M_\odot)^{2.3}$ for masses less than 0.43 solar. The pressure wave amplitude is then set to 20 mbar at the Earth-equivalent insolation, i.e. $\delta p = 20 \left(a_{p-equiv}/a_p\right)^{-2}$ mbar. Using 20 mbar as a fiducial amplitude for the pressure wave near resonance should be considered a very basic approximation. However, the linear dependency of $T_{thermal}$ on $\delta p$ suggests that current uncertainty in this amplitude is not the largest factor that might impact the conclusions presented here.

Despite the many parameter uncertainties (from phase lag angles in both torques, atmospheric properties, and the true range of planetary sizes to be considered), it is apparent from Fig 2. that a wide variety of circumstances could lead to conditions where thermal and stellar torques are comparable and planet rotational evolution in the liquid-water orbital zone could stall long before synchronous conditions are approached. This is especially true for stellar masses larger than $0.3 M_\odot$, where modest variations in tidal bulge phase lags (e.g. Correia & Laskar 2010) can easily bring the opposing torques amplitudes into a range where balance can occur. Furthermore, for a true Earth-analog with a $1 M_\odot$ host star, but no large moon in a lunar configuration, it appears possible for the near-resonance thermal tide torque to significantly impact the long term rotational evolution.

**2.1 Implications for Earth-analogs**

The amplitude of thermal tide resonant torques suggests that an unbiased sample population of atmosphere-bearing, Earth-analog planets around lower mass stars might bifurcate between near-synchronous rotation and rotation much closer to the terrestrial $\sim 21\ hr$ resonance daylength (modulo the system age and details such as surface temperature and composition). An Earth-analog world orbiting a $0.3\ M_\odot$ star in the habitable zone has a nominal 'spin-down' time to a synchronous state of approximately $10^7 - 10^8\ yr$ considering solar tidal torques alone and assuming an initial rotation period of $\sim 12\ hr$ (e.g. Peale 1977, Gladman *et al.* 1996). By comparison, the same planet orbiting a 0.5 $M_\odot$ star would reach synchronicity after approximately $10^9\ yr$. Therefore, in this stellar mass range, resonance trapping could occur in as little as $\sim 10^7 - 10^8 \ yr$ after the final epoch of planet formation.

That trapping has the potential to influence both the early and long-term climatic states and potential early biosphere (based on the example of Earth's history), together with their feedback on the thermal tide conditions themselves. Global mean surface temperature directly influences the daylength of resonance (see above, by altering the atmospheric scale height). Warmer planets, towards the inner edge of their surface-liquid-water orbital zones, should have systematically shorter resonant daylengths (e.g. Lindzen & Chapman 1971, Bartlett & Stevenson 2016, according to the $\sqrt{m/T}$ scaling), with colder planets expected to have longer resonance daylengths. In both cases the variance is at the level of $\sim 0.5 - 1.0 \ hr$.

Furthermore, faster rotating Earth-analog planets are expected to have less efficient latitudinal energy transport (e.g. Spiegel *et al.* 2008). This can lead to hotter equators and colder high latitudes (for low obliquities) and can reduce the net liquid-water surface area of the planet if all other properties remain the same. For rotation variations of the order 10% (e.g. around resonance in this present study) the effect will be modest, but is possibly relevant for biosphere productivity. But for an Earth-analog world around a low mass star where the timescale to reach a slow rotation rate is only $\sim 10^7 - 10^9 \ yr$, a stall that maintains 'fast' rotation for even a few tens of millions of years could have a major effect on environmental history and any nascent biosphere.

There is also evidence from more sophisticated three-dimensional climate modeling (Way *et al.* 2015 and refs. therein) that slow-rotating Earth-analog planets may have systematically lower global mean temperature for a fixed stellar input, due to albedo changes from cloud formations in the substellar region. Therefore, two populations of otherwise similar planets, one 'stalled' at faster rotation, one close to synchronous rotation, could exhibit systematically different global mean temperatures for similar stellar input.

**2.2 Biosphere influence**

Oxygenation of an atmosphere, and/or the formation of a significant ozone layer can alter the resonant daylength for thermal tides by $\sim 0.5\ hr$ (Lindzen & Chapman 1971). Ozone mediated stratospheric heating shortens the resonant daylength. Higher $O_2$ abundance lengthens the daylength by decreasing the atmospheric scale height – assuming displacement of an $N_2$ rich atmosphere (e.g. Zahnle & Walker 1987). However, ozone enhancement would be expected early in an oxygenation event, before any significant rise in $O_2$ (e.g. Kasting 1985).

On the modern Earth, ozone, and its production, plays a key role in increasing the efficiency of atmospheric excitation – by absorbing sunlight over tens of kilometers in the 'middle' of the atmosphere (10-50 km), spanned by the vertical wavelength of the semidiurnal thermal tide. Therefore, in addition to the resonant period, the *amplitude* of the thermal tide must be also strongly dependent on the efficiency of absorption and the molecular culprits. Both ozone and methane are high efficiency absorbers.

This raises the intriguing possibility that the rotational evolution of Earth-analog planets can be influenced by biological activity. Three interesting scenarios are: (1) a rise in ozone prior to a planet slowing to the resonant state, (2) a rise in ozone during resonance, and (3) a buildup of $O_2$ during and immediately after resonance. In (1) the shortened daylength of resonance could cause rotation stalling at earlier times. In scenario (2) the change in resonance conditions would shift the curve in Fig. 1 to the right, and could therefore break the torque balance (c.f. Zahnle & Walker 1987), releasing the planet to its usual spin-down state. In the case of (3), depending on the precise timing, and assuming rotational stalling, the resonant conditions could be shifted 'ahead' of the rotational slowdown – either causing the planet rotation evolution to re-stall in the future or simply extending the time spent in its already stalled state. In all cases, ozone's impact on the resonant tide amplitude could act to increase the importance of these effects.

Evidence from Earth's oxygenation also indicates that the chemical changes induced on the atmospheric greenhouse via the breakdown of methane could have caused a precipitous drop in global temperature (e.g. Kopp *et al.* 2005, Frei *et al.* 2009). Global temperature drops of $\sim 20 - 30\ K$ could increase the daylength of thermal tide resonance by $\sim 1 - 2\ hr$. If analogous events occur during scenario (3) it is conceivable that a planet could experience more than one rotational stalling, one during oxygenation and one later during the induced 'snowball' event.

Clearly the frequency with which such coincidences might happen on Earth-analog worlds around low-mass stars is extremely hard to quantify. Nonetheless, the relevant tidal evolution timescales around lower-mass stars are closely matched to the timescales of early terrestrial biosphere innovations (photosynthesis, eukaryotic cells). Given the theoretical evidence for the robustness of resonant thermal and solar torque balances once they are established (Bartlett & Stevenson 2016) it can be hypothesized that biospheres might experience and even help sustain very long periods of essentially fixed, comparatively fast, planetary rotation, rather than the typically assumed slow synchronous rotation.

Although an extreme extrapolation, this discussion raises the question of whether there exist any conceivable forms of active feedback between planetary rotation rate and the propensity of a biosphere for producing an oxygen-rich environment. (A similar discussion could be made for other species that strongly effect atmospheric absorption of stellar radiation, e.g. methane). Slowing rotation due to stellar tides (or natural satellites) will induce climate change, at a significant rate for Earth-analog worlds around low-mass stars due to the shorter timescale of tidal evolution. If a specific planetary environment is more conducive to selective pressure favoring a process such as oxygenesis (e.g. faster rotating, regular illumination of entire surface) then any feedback sustaining the environment during the establishment of the biological mechanism would presumably be favored.

## 3. Conclusions

To properly assess the likelihood of thermal tide resonance and rotational torque balance in Earth-analog worlds, many details need to be studied. Further insight to thermal tides close to resonance could be accomplished with three-dimensional general circulation models (GCMs, e.g. Way *et al.* 2017), coupled to a dynamic model of planetary torques and spin-orbit evolution. Investigating the interplay of different atmospheric species and their effect on resonance periods and pressure amplitudes could be informative, especially if species have long-term bio-geophysical impact (e.g. $N_2$ and ammonium subduction, $CH_4$, $CO_2$).

Although extremely challenging (e.g. Spiegel *et al.* 2007), future observational constraints on rocky planet rotation rates and their population statistics could also provide clues to atmospheric evolution and the possibility of biological influence.

**Acknowledgments**

This research was supported by the NASA Astrobiology Program through participation in the Nexus for Exoplanet System Science and NASA grant NNX15AK95G. The author thanks Norm Sleep and an anonymous reviewer for comments that have significantly improved the manuscript.

**Author Disclosure Statement**

No competing financial interests exist.

**FIGURES**

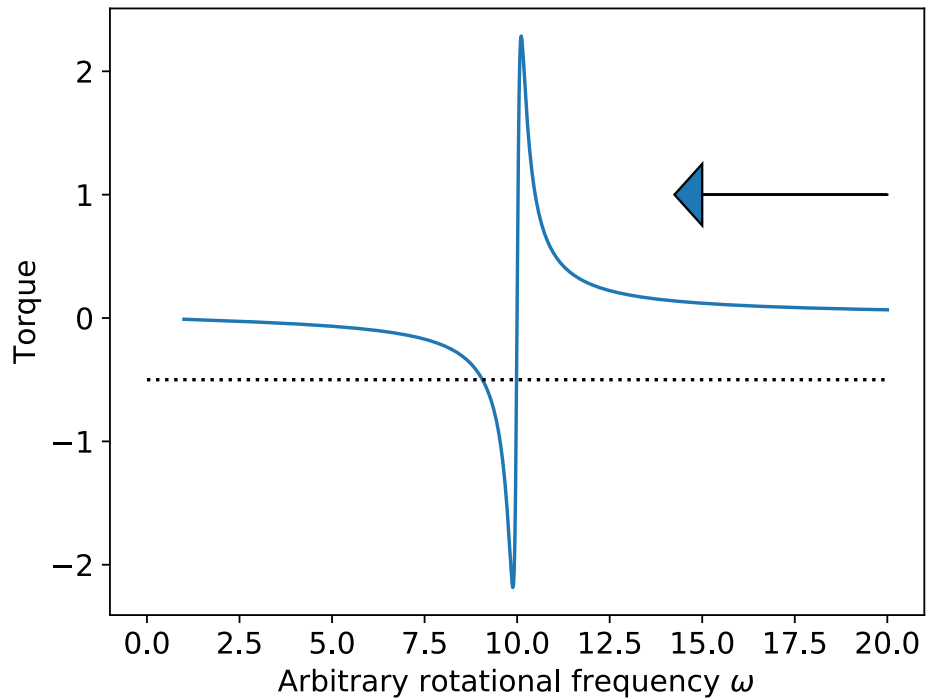

**FIG. 1.** The basic form of opposing thermal tide torque (plotted curve) during passage through resonance (e.g. Zahnle & Walker 1987, Bartlett & Stevenson 2016). Arrow indicates approach from higher rotational rate. Dotted line indicates a hypothetical decelerating oceanic/body torque. In this case, the accelerating (opposing) thermal tide torque can balance the decelerating torque at two closely spaced rotational frequencies, but only the interior point at higher frequency is a stable equilibrium.

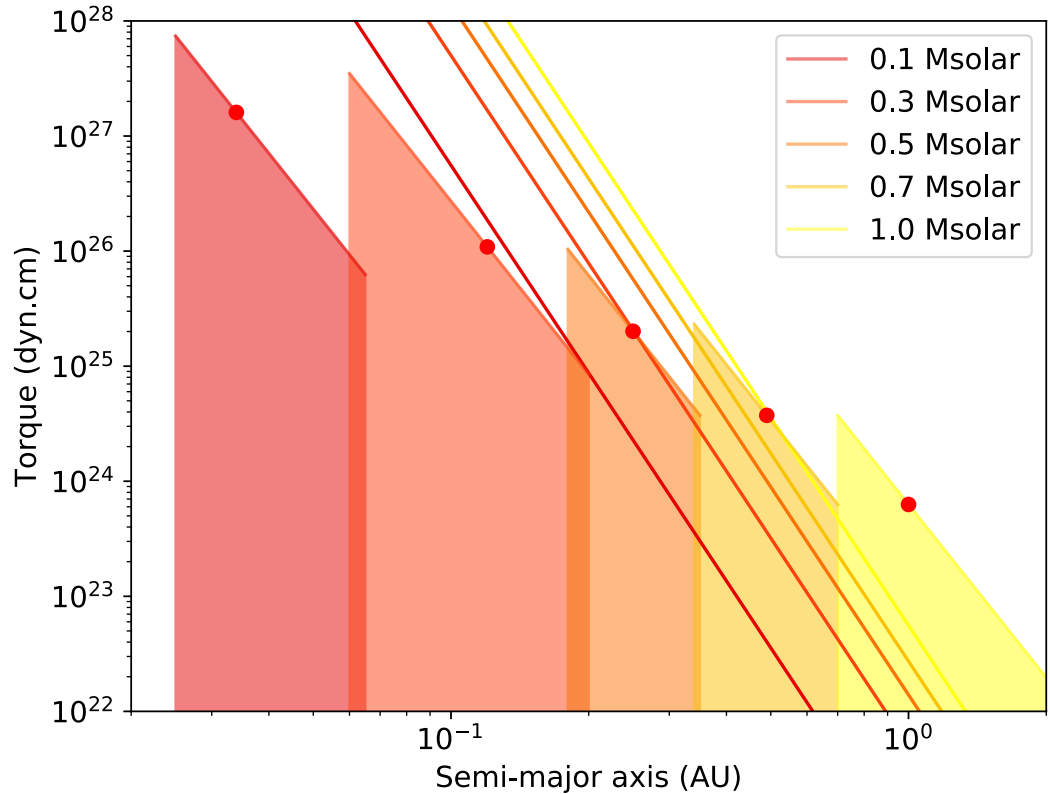

**FIG. 2.** Uppermost curves correspond to maximum stellar tidal torques for varying stellar masses (see legend) versus orbital semi-major axes. Filled regions indicate range of potential resonant state thermal tidal torques for Earth-analog planets. Liquid water orbital ranges compiled from various sources as: 0.025-0.065 AU, 0.06-0.2 AU, 0.18-0.35 AU, 0.34-0.7 AU, and 0.7-2.0 AU for stellar masses of 0.1, 0.3, 0.5, 0.7, and 1.0 $M_\odot$ respectively (see e.g. Cockell *et al.* 2016). Solid circle markers indicate the locations of Earth-equivalent stellar radiation input (assuming modern solar luminosity).